\documentclass[review]{elsarticle}

\journal{Journal of Optical Society of America - B}

\begin{document}

\begin{frontmatter}

\title{Selecting the pre-detection characteristics for effective fiber coupling of entangled photon sources}

%\author{Ali Anwar\fnref{myfootnote}}
\author[1]{Ali Anwar}
\author[1]{Chithrabhanu P}
\author[1]{Salla Gangi Reddy}
\author[1,2]{Nijil Lal}
\author[1]{R. P. Singh}
\address[1]{Physical Research Laboratory, Navarangpura, Ahmedabad, India-380009.}
\address[2]{IIT Gandhinagar, Palaj, Gandhinagar, India-382355.}

\cortext[mycorrespondingauthor]{Corresponding author: alianwar@prl.res.in (Ali Anwar)}

\begin{abstract}
Photon modes have an important role in characterizing the quantum sources of light. Proper coupling of various photon modes obtained in spontaneous parametric down conversion (SPDC) process in optical fibers is essential to generate an effective source of entangled photons. The two main pre-detection factors affecting the biphoton mode coupling in SPDC are the pump beam focusing parameter and the crystal thickness. We present the numerical and experimental results on the effect of pump focusing on conditional down-converted photon modes for a Type-I BBO crystal. We experimentally verify that biphoton coupling efficiency decreases asymptotically with pump beam focusing parameter. We attribute this behaviour to (a) the asymmetry in the spatial distribution of down-converted photons with the pump beam focusing parameter and (b) the ellipticity of biphoton modes introduced due to the focusing of the pump beam. We also show experimentally the ellipticity as well as quantify the asymmetry with the pump focusing parameter. These results may be useful in selecting optimum conditions in the down-conversion process for generating efficient sources of  entangled photons for quantum information applications.
\end{abstract}


\end{frontmatter}

%\linenumbers

\section{Introduction}

\par Spontaneous parametric down conversion is one of the most popular methods used to generate heralded single photons as well as entangled photon pairs \cite{burnham}. In this second order non-linear optical process, a pump photon of higher energy, when interacting in a non-linear medium, down-converts to two lower energy photons. It is governed by the conservation laws of energy and momentum. The momentum conservation is apparently called as phase matching condition \cite{boydbook}. Photon pairs that are entangled in different degrees of freedom such as polarization \cite{kwiat,eibl,zhi} and orbital angular momentum \cite{mair} are experimentally realized using SPDC process. In recent years, multiple SPDC setups are extensively used for generating multipartite entangled states \cite{lu,hamel}. Due to applications in quantum teleportation \cite{bennett}, quantum cryptography \cite{ekert}, quantum metrology \cite{bell} and super dense coding \cite{bennett2}, there is an increased demand of efficient single photon and entangled photon sources. 

\par There have been many theoretical studies on effective fiber coupling of SPDC sources of photons \cite{bovino,castelletto,castellettoconf,dragan,andrews,daniel,fedrizzi,bennink}. The main parameters that control the collection efficiency of photon pairs are thickness of the crystal used for down-conversion, spatial walk-off, and mode field diameter of the optical fibers effectively imaged onto the crystal plane \cite{bovino}. Coupling of SPDC photons with single-mode and multi-mode fibers were investigated \cite{castelletto,castellettoconf} as a function of pump beam diameter, crystal thickness and walk-off. Significant changes were observed in the coupling of SPDC photon pairs to the single-mode and multi-mode fibers while varying the pump beam diameter. It was analytically shown that the coincidence spectrum becomes inseparable under strong pump focusing conditions \cite{dragan}. It has been claimed that the important parameters for mode coupling in collinear parametric down conversion are the photon wavelength, the focal length of lens and the fiber diameter \cite{andrews}. Similar work has been carried out for quasi-phase matched crystals, but coupling was investigated by considering the down-converted output as a classical beam \cite{daniel}. Dependence of photon coupling ratio on focusing parameter of pump and collection modes, and the crystal length in the case of periodically-poled crystals has been studied \cite{fedrizzi} with an emphasis on grating defects. The theoretical framework of Bennink \cite{bennink} shows that the optimum focusing conditions for maximum efficiency of collinear PDC are precisely same as that of sum frequency generation and parametric amplification using Gaussian beams. On the contrary, Smirr \textit{et. al.} have shown that there is no significant change in the coupling efficiency of conditional biphoton modes with the focusing parameter, under collinear phase-matching conditions \cite{jean,boydpaper}.

\par Here, we study the effect of pump focusing on biphoton coupling efficiency of photon pairs obtained in a non-collinear SPDC process experimentally. We observe that the coupling efficiency decreases asymptotically with the focusing parameter of the pump beam. We give theoretical explanation on how crystal thickness influences the behaviour of biphoton modes in pump focusing. We also give a physical reason for this decrease in coupling efficiency using the matching of conditional optical modes of down-converted photons, which has not been addressed earlier. We show that a \textit{loosely focused pump beam and a thin crystal} are the best pre-detection conditions for the effective fiber coupling of entangled photons, as the former reduces the effect of SPDC ring asymmetry and the later reduces the walk-off effects inside the crystal. We also verify that the role of collection mode diameter on mode coupling to the fibers is more significant in tight pump focusing than in loose pump focusing.

\section{Theory of SPDC: Semi-classical treatment}
In the multi-mode perturbative treatment of SPDC, we consider the electric field of a pump beam as classical and the down-converted photons as quantum. Their respective electric fields can be written as \cite{ling}

\begin{equation}
E_p(\mathbf{r}, t)=\frac{1}{2}[\alpha_p\mathbf{e}_p E_p^0g_p(\mathbf{r})e^{-i\omega_pt}+c.c]
\end{equation}\\
\begin{equation}
\hat{E}_{s,i}(\mathbf{r}, t)=\frac{i}{2}\sum_{k_{s,i}}\alpha_{s,i}\mathbf{e}_{s,i}\sqrt{\frac{\hbar\omega_{s,i}}{2n_{s,i}^2\epsilon_0V_Q}}g_{s,i}(\mathbf{r})e^{-i\omega_{s,i}t}\hat{a}_{k_{s,i}}(t)+h.c
\end{equation}\\
where $\alpha_j=\sqrt{2/\pi w_j^2}$, ($j=p,s,i$), $w_j$ is the beam waist, $\mathbf{e}_p$ is the polarization vector of pump, $\hat{a}_{k_{s,i}}(t)$ is the annihilation operator, $E_p^0$ is the amplitude of the pump beam and $V_Q$ is the quantization volume. $\mathbf{e}_{s,i}$, $n_{s,i}$ and $\omega_{s,i}$ are respectively the polarization vectors, refractive indices and angular frequencies  of target modes ($s$-signal, $i$-idler).  $g_j(\mathbf{r})$ is the spatial mode function for the electric field and $\epsilon_0$ the dielectric constant. Here, $c.c$ and $h.c$ represents the complex conjugate and the hermitian conjugate respectively. The Hamiltonian governing the interaction of pump with the non-linear crystal is given by \cite{ou}
\begin{equation}
\hat{H}_I(t)=\int 2\epsilon_0\left(\chi^{(2)}(\mathbf{r}):E_p^{(+)}(\mathbf{r}, t)E_s^{(-)}(\mathbf{r}, t)E_i^{(-)}(\mathbf{r}, t)+h.c\right)d^3\mathbf{r}
\end{equation}\\
where $E_p^{(+)}(\mathbf{r}, t)$ is the positive frequency part of the pump field, $E_s^{(-)}(\mathbf{r}, t)$ \& $E_i^{(-)}(\mathbf{r}, t)$ are the negative frequency parts of signal \& idler modes respectively. $\textbf{r}=(x, y, z)$ is the position vector in spatial coordinate system, $\chi^{(2)}(\textbf{r})$ is the non-linear susceptibility tensor. For an interaction time $\tau$, the quantum state of photon pairs generated in the process can be obtained by applying a time evolution operator $\exp[(-i/\hbar)\int_{0}^{\tau}H_I(t)dt]$ on the initial state. By truncating the above exponential function to the first order, the state is given by

\begin{equation}
|\Psi(t)\rangle \approx \left(1-\frac{i}{\hbar}\int_{0}^{\tau}H_I(t)dt\right)|\Psi(0)\rangle
\label{qtmstate}
\end{equation}\\
where $|\Psi(0)\rangle$ is the initial joint state of signal and idler. For a spontaneous process, the initial states are the vacuum states in the momentum space.

\par In general, each parametric down conversion process can be characterized by a joint two-photon mode function  of momentum and frequency, which is derived from the quantum state of down-converted output as given in Eqn. \ref{qtmstate} \cite{clara}. The mode function gives the information about the process such as pump beam characteristics and crystal phase matching conditions. Using mode function, we can quantify spatial and spatio-temporal correlations among the down-converted modes without actually doing the state tomography \cite{clara}. The mode function has a one-to-one correspondence with the coincidence counts that we measure in experiment. A typical two-photon mode function \cite{terriza} in transverse momentum coordinates $\mathbf{k}_s^{\perp}$ \& $\mathbf{k}_i^{\perp}$ is given by

\begin{equation}
\Phi(\mathbf{k}_s^{\perp},\mathbf{k}_i^{\perp},\Delta k)=E_0(\mathbf{k}_s^{\perp}+\mathbf{k}_i^{\perp})sinc\left(\frac{\Delta kL}{2}\right)exp\left(i\frac{\Delta kL}{2}\right)
\label{modefna}
\end{equation}\\
where $E_0(\mathbf{k}_s^{\perp}+\mathbf{k}_i^{\perp})$ represents the pump transverse wave vector amplitude distribution, $\Delta k$ is the phase mismatch, and $L$ is the thickness of the crystal. The exponential factor in the Eqn. \ref{modefna} is a global phase term.

\section{Mode coupling techniques in SPDC}
Better sources of single photons by parametric down-conversion require the efficient coupling of optical modes involved in the process, into the fiber. Before coupling, the down-converted photons are spatially and spectrally filtered. The functions that represent the spatial and frequency filtering ($\omega_c$) of down-converted photons are given by

\begin{equation}
\Gamma_{\textit{spatial}}=\exp\left(-\frac{w_c^2}{2}|\mathbf{k}_c^2|\right)
\end{equation}

\begin{equation}
\Gamma_{\textit{frequency}}=\exp\left(-\frac{(\omega_c-\omega_{c0})^2}{2B_c^2}\right)
\end{equation}
where $w_c$ and $\mathbf{k}_c$ are respectively the spatial collection mode width and the transverse momentum coordinate of collection mode. $\omega_{c0}$ and $B_c$ are the central wavelength and the bandwidth of the frequency filter respectively. In biphoton mode coupling, first we define a reference mode by imaging the single mode fiber-coupled idler photons onto the crystal. i.e., we project them into a single mode Gaussian.

\begin{equation}
u(k_i^{\perp})=\exp\left(-\frac{w_i^2}{4}|\mathbf{k}_i^{\perp}|^2\right)
\end{equation}
where $\mathbf{k}_i^{\perp}$ and $w_i$ are respectively the transverse wave-vector and diameter of the idler mode. The conditional angular distribution of the down-converted light in the signal arm has to be matched with this reference mode. Thus, the spatial distribution of signal photon is given by the normalized mode function

\begin{equation}
\phi_s(\mathbf{k}_s^{\perp},\Delta k)=N_s\int d\mathbf{k}_i^{\perp}\Phi(\mathbf{k}_s^{\perp},\mathbf{k}_i^{\perp},\Delta k)u(\mathbf{k}_i^{\perp})
\label{condmodefn}
\end{equation}
where $N_s$ is the normalization factor. The conditioned spatial distribution of the idler photon, $\phi_i(\mathbf{k}_i^{\perp},\Delta k)$, can also be obtained using similar calculations. The overlap of these two conditioned modes is what we call mode matching. The amount of mode matching \cite{aichele} between conditional modes of signal and idler is given in terms of their respective mode functions as

\begin{equation}
M_{si}=\frac{\int d^3\mathbf{k}_sd^3\mathbf{k}_i\phi_s(\mathbf{k}_s^{\perp},\Delta k)\phi_i^*(\mathbf{k}_i^{\perp},\Delta k)}{\left[\int d^3\mathbf{k}_s\phi_s(\mathbf{k}_s^{\perp},\Delta k)\right]\left[\int d^3\mathbf{k}_i\phi_i(\mathbf{k}_i^{\perp},\Delta k)\right]}
\end{equation}
Purity of each mode $j$ ($j=s,i$) is given by

\begin{equation}
P_j=\frac{\int d^3\mathbf{k}_jd^3\mathbf{k}_j'\phi_j(\mathbf{k}_j^{\perp},\Delta k)\phi_j^*(\mathbf{k}_j^{'\perp},\Delta k')}{\left[\int d^3\mathbf{k}_j\phi_j(\mathbf{k}_j^{\perp},\Delta k)\right]^2}
\end{equation}
where $\mathbf{k}_j^{\perp}\neq\mathbf{k}_j^{'\perp}$. By simple manipulation using Cauchy-Schwarz inequality, we obtain the criteria for optimal mode matching of biphoton modes in the SPDC as
\begin{equation}
M_{si}^2\leq P_sP_i
\end{equation}
where the equality is achieved for perfect mode matching.

\section{SPDC with focused pump}
\begin{figure}[h]
\begin{center}
\includegraphics[width=0.8\textwidth]{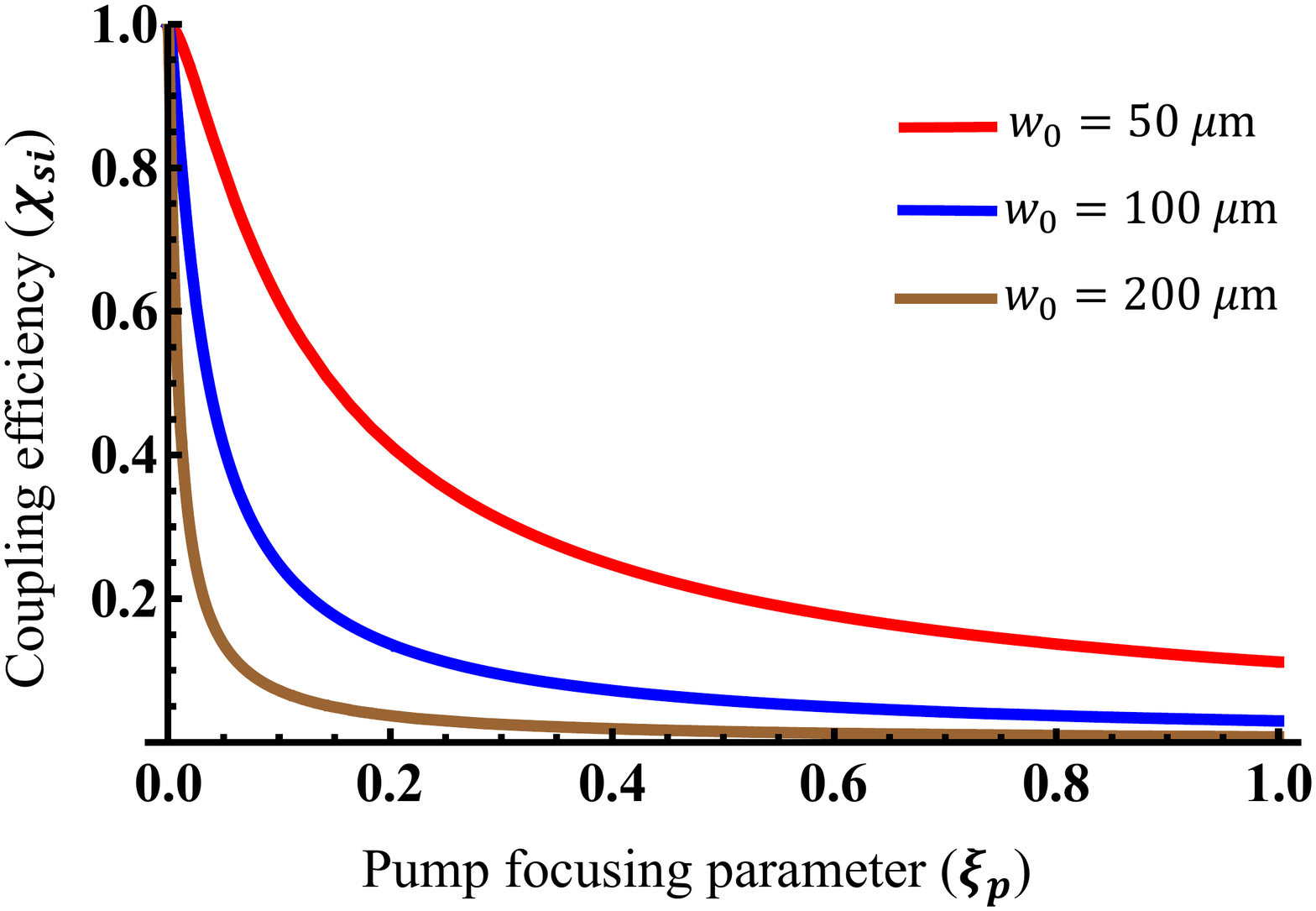}
\caption{Numerical plot of mode coupling efficiency vs pump focusing parameter with different collection mode diameters.}\label{theoryplot}
\end{center}
\end{figure}

Here we study the effect of pump focusing on mode coupling efficiency. A focused Gaussian pump beam is characterized by a focusing parameter $\xi_p$ \cite{boydpaper}, given by

\begin{equation}
\xi_p=\frac{L}{k_pw_p^2}
\end{equation}
where $k_p$ is the magnitude of wave vector for the pump beam, $L$ is the crystal thickness and $w_p$ is the pump beam waist located inside the crystal. A field-based photon collection efficiency is given by
\begin{equation}
\chi_{si}=\frac{C_{si}}{\sqrt{C_sC_i}}
\label{photonpair}
\end{equation}\\
where $C_{si}$ is the measure of overlap of all the three modes - the correlated biphoton mode with the signal mode and the idler mode. While  $C_s$ \& $C_i$ are the measures of overlap between the correlated biphoton mode and a single mode. In experiment, $C_{si}$ turns out to be the coincidence counts and $C_s$, $C_i$ are the singles count of signal and idler respectively. In terms of focusing parameters of pump and the diameter of the collection (signal \& idler) modes, the coupling efficiency for a degenerate parametric down conversion under thin crystal approximation can be rewritten as \cite{castelletto}
\begin{equation}
\chi_{si}=\frac{4L(k_p\xi_pw_0^2+L)}{(k_p\xi_pw_0^2+2L)^2}
\end{equation}\\
where $k_p$ is the magnitude of wave vector of the pump mode and $w_0$ is the diameter of each target modes. The numerical plot of mode coupling efficiency vs pump focusing for different collection mode diameters ($w_0$) is given in Fig. \ref{theoryplot}. For a fixed value of focusing parameter, the coupling efficiency is more for smaller collection mode diameter. This behaviour is more pronounced in the loose focusing region ($\xi_p$ < 0.1).

A crude estimation of the effective crystal length ($L_{eff}$) along which down-conversion takes place is  given in \cite{vicent}. This effective crystal length depends on the pump focusing parameter, the orientation of propagation vectors of the down-converted photons with respect to the pump beam, and the collection mode diameter ($w_0$). This is used to distinguish the behaviour of biphoton mode in different crystal length regimes \cite{ramirez}. In short crystal regime ($L$ < $L_{eff}$), the biphoton mode is completely determined by the pump properties, i.e. the crystal length effects can be neglected. In the long crystal regime ($L$ > $L_{eff}$), the biphoton mode depends on the pump as well as the crystal properties.

\section{Experimental details}
\begin{figure}[h]
\begin{center}
\includegraphics[width=1\textwidth]{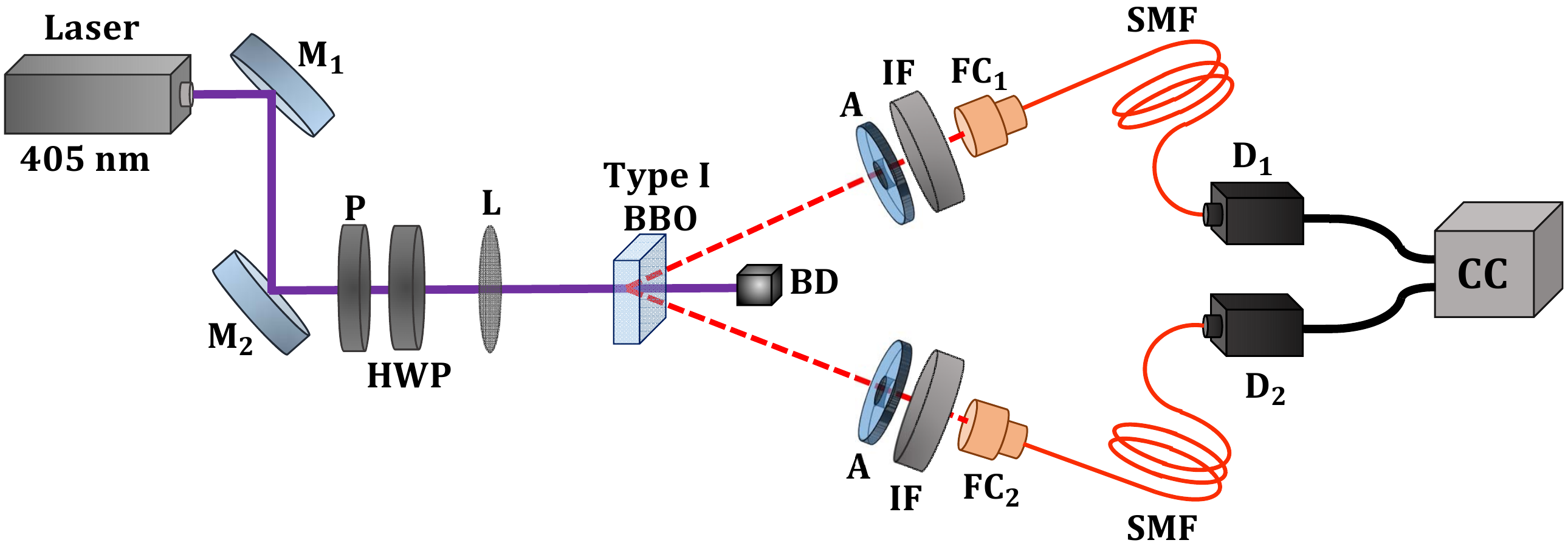}
\caption{Experimental setup used for generating the correlated photon pairs through SPDC process. Here, M$_1$,M$_2$ - Plane mirrors; P - Polarizer; HWP - Half Wave Plate; L - Plano convex lens; A - Aperture; IF - Interference filter; BD - Beam dump; FC$_1$,FC$_2$ - Fiber collimators; SMF - Single Mode Fiber; D$_1$, D$_2$ - Single photon counting modules(SPCM's); CC - Coincidence Counter.}\label{exptsetup}
\end{center}
\end{figure}

The experimental setup used to verify the above theoretical arguments is given in Fig. \ref{exptsetup}. Here, we have used a UV diode laser (Toptica iBeam smart) of wavelength 405 nm and power 300 mW with a spectral band width of 2 nm, to pump the non-linear crystal, Type-I $\beta$-Barium Borate (BBO), of thickness 2 mm and transverse dimensions of 6 mm$\times$6 mm with an optic axis oriented at 29.97$^\circ$ to the normal incidence. The combination of a polarizer and a half wave plate allows us to vary the pump beam polarization along the crystal axis. A plano-convex lens is used to focus the pump beam inside the crystal kept at the focal plane. For changing the focusing parameters, plano-convex lenses of focal lengths 50, 100, 150, 200, 250, 300 \& 750 mm are used in our experiment. The down-converted photons (signal \& idler) of wavelength 810 nm each are generated in a non-collinear fashion at diametrically opposite points of the SPDC ring. The images of the down converted ring for different focusing conditions were taken using an Electron Multiplying CCD (EMCCD) camera (Andor iXon3) of 512$\times$512 pixels with a pixel size of 16$\times$16 $\mu m^2$. 

To measure the number of photon pairs generated, two diametrically opposite portions of the SPDC ring at a given plane were selected using apertures (A) and the photons coming out of each aperture were collected using the fiber collimators FC$_1$ \& FC$_2$ (CFC-5X-B, Thorlabs) of focal lengths 4.6 mm and a numerical aperture of 0.53. The fiber collimators are attached to the single mode fibers (P1-780A-FC-2, Thorlabs) each having a numerical aperture of 0.13 and a mode field diameter of 5$\pm$0.5 $\mu$m, which inturn are connected to the single photon detectors D$_1$ \& D$_2$ (SPCM-AQRH-16-FC, Excelitas). The detectors have a timing resolution of 350 ps with 25 dark counts per second. Two interference filters (IF) of passband 810$\pm$5 nm are kept very close to the fiber collimators to make sure that other unwanted wavelengths are properly filtered out. To measure the number of correlated photon pairs, the two detectors are connected to a coincidence counter (CC), IDQuantique-ID800, having a time resolution of 81 ps.

\section{Results and Discussion}
To study the effect of pump focusing on biphoton modes, first we study the effect of asymmetry present in the angular distribution of down-converted photons obtained for different focusing parameters. For this, we image the ring of down-converted photons using an Electron Multiplying CCD camera.
\begin{figure}[h]
\begin{center}
\includegraphics[width=0.7\textwidth]{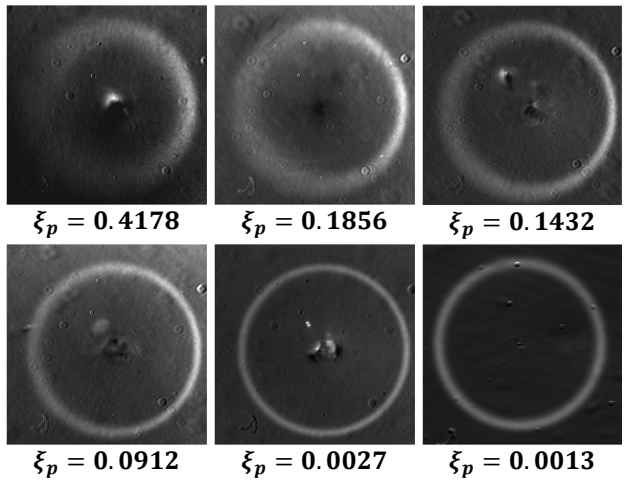}
\caption{Electron Multiplying CCD images of the down converted rings for different pump focusing parameters ($\xi_p$) corresponding to focal lengths \textit{f}=100, 150, 200, 300, 600 \& 750 mm respectively.}\label{emccdimages}
\end{center}
\end{figure}
These images are shown in Fig. \ref{emccdimages} corresponding to the different pump focusing parameters obtained by focusing the pump beam on to the crystal using different plano-convex lenses of focal lengths 100, 150, 200, 300, 600 \& 750 mm. The inhomogeneity in the spatial distribution of the down-converted photons decreases with decrease in the pump focusing parameter, i.e. with the decrease in the focal length of the lens used to focus the pump beam.

\begin{figure}[h]
\begin{center}
\includegraphics[width=0.7\textwidth]{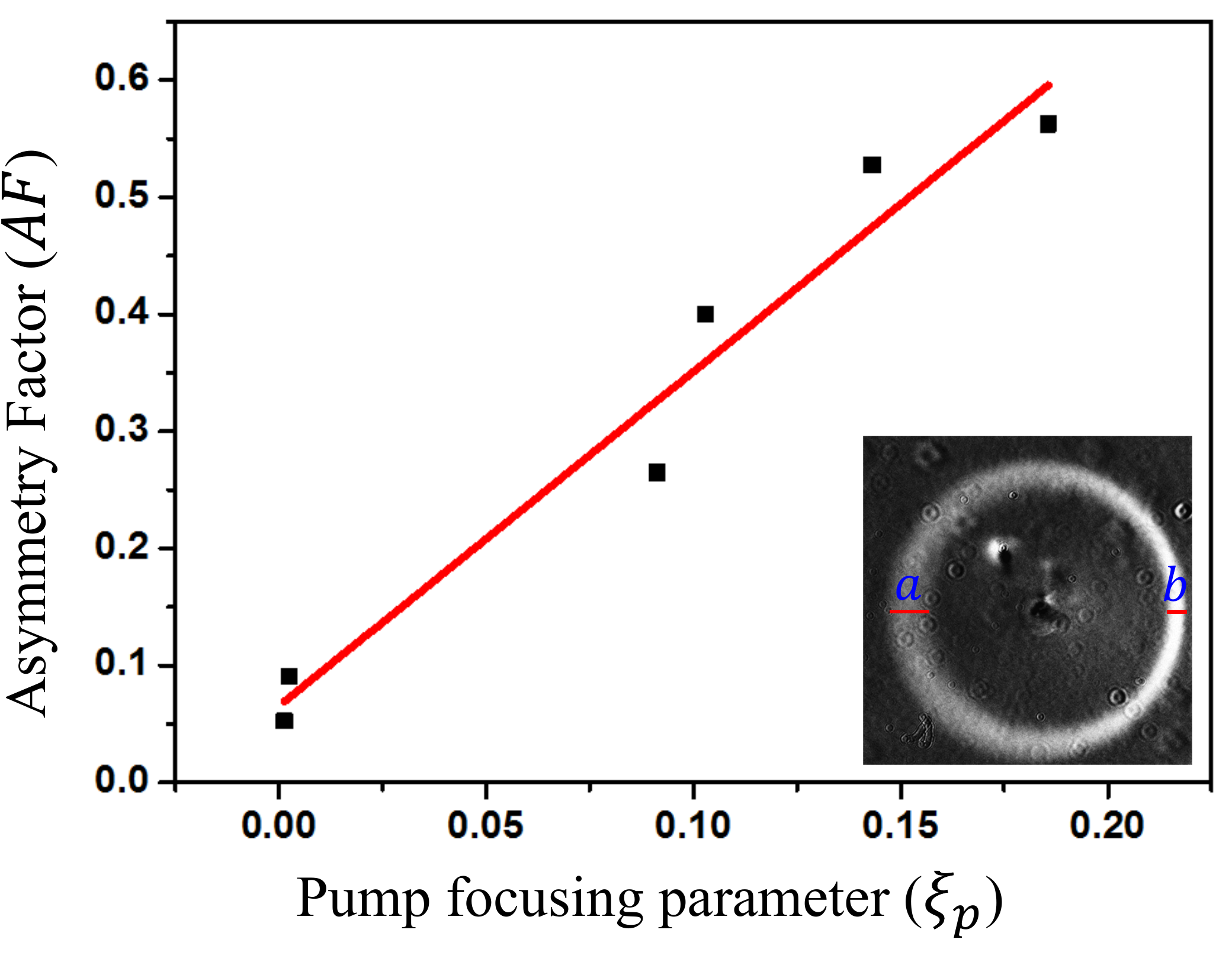}
\caption{(Colour online) Variation of SPDC ring asymmetry with pump beam focusing.}\label{asymmetryvsfocus}
\end{center}
\end{figure}

In order to quantify the asymmetry of a ring formed by the down-converted photons, we introduce an asymmetry factor (\textit{AF}), which is defined as

\begin{equation}
AF=1-\frac{b}{a}
\end{equation}
where $a$ and $b$ are the ring widths at two diametrically opposite ends of the down-converted ring ($a$ > $b$) as shown in the inset of Fig. \ref{asymmetryvsfocus}.
The asymmetry factor is calculated for different focusing parameters of the pump beam and shown in Fig. \ref{asymmetryvsfocus}. We found that the asymmetry increases with the increase in focusing parameter i.e. when we move from loose focusing to tight focusing of the pump beam. Also, we observed that the asymmetry is independent of the propagation of down-converted photons from the crystal plane and depends on the crystal tilt in a known fashion \cite{ramirez}. The variation of asymmetry factor is linear with respect to the focusing parameter. To study the influence of the crystal chosen ($L$=2 mm), we calculated the effective crystal length ($L_{eff}$) \cite{vicent} for each values of pump focusing parameter used in the experiment. Under tight focusing condition ($\xi_p > 0.1$), $L_{eff}$ is $\sim$12.3 mm and for loose focusing condition ($\xi_p < 0.1$), it is $\sim$13.9 mm. So, the range of pump focusing parameters we considered in the experiment were found to satisfy the condition $L < L_{eff}$ \cite{ramirez}, i.e. the influence of crystal length on the asymmetry in the SPDC ring is negligible when compared with that of the pump focusing parameter.

\begin{figure}[h]
\begin{center}
\includegraphics[width=0.4\textwidth]{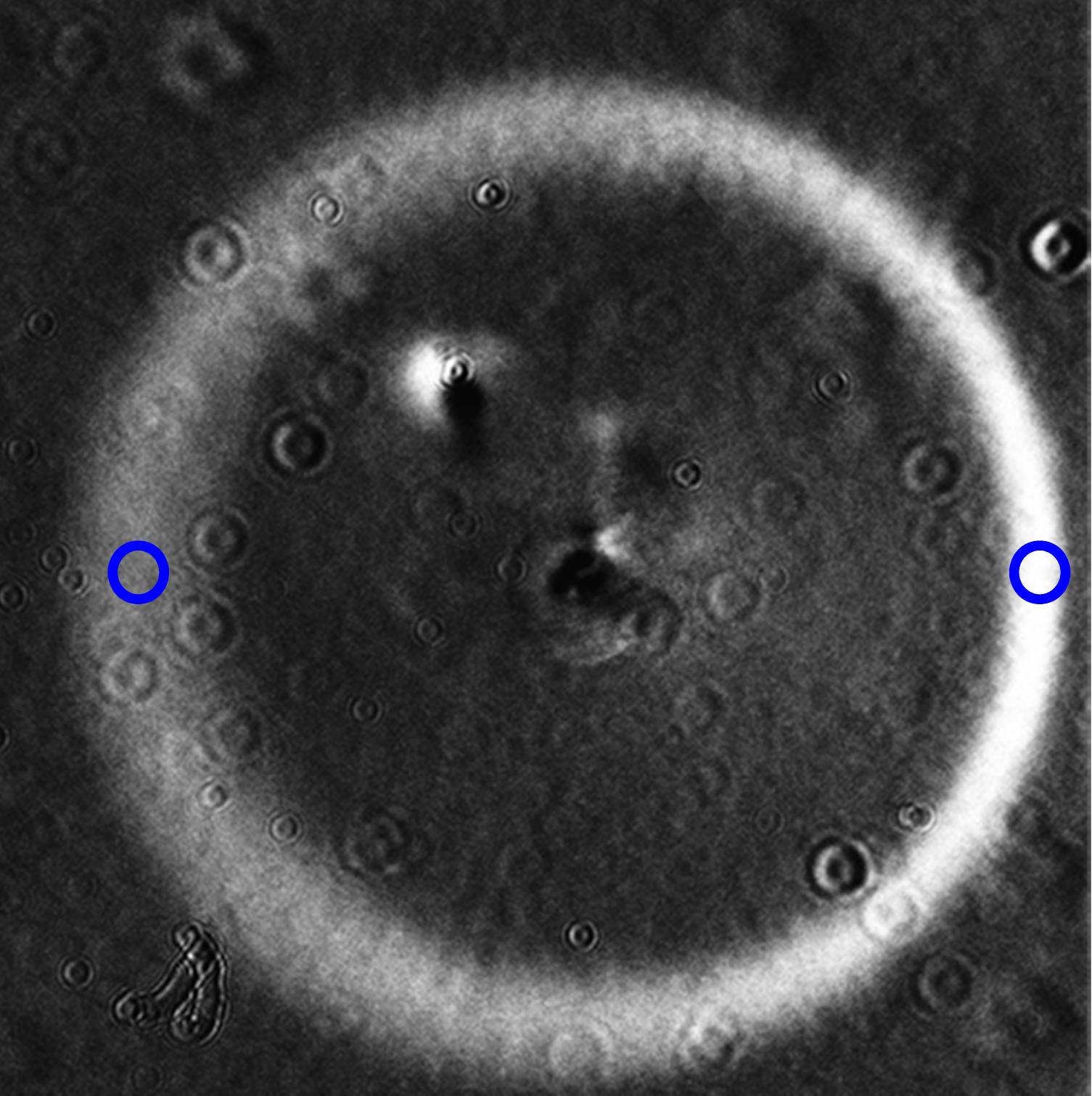}
\caption{(Colour online) EMCCD image of a ring of down-converted photons. For coincidence counting setup, the two diametrically opposite points of the ring (shown in blue circles) are selected.}\label{emccdimageselectaperture}
\end{center}
\end{figure}

For coincidence detection, we choose two diametrically opposite portions of the down-converted ring using two apertures of same width as shown in Fig. \ref{emccdimageselectaperture}. Because of the asymmetry of SPDC ring, the photon number densities (number of photons per unit area) of signal and idler in the selected areas are different due to which we are not able to select all the signal photons that correspond to the selected idler photons. This accounts for the asymptotic decrease in mode coupling efficiency of down-converted photon pairs with pump focusing.

\begin{figure}[h]
\begin{center}
\includegraphics[width=0.7\textwidth]{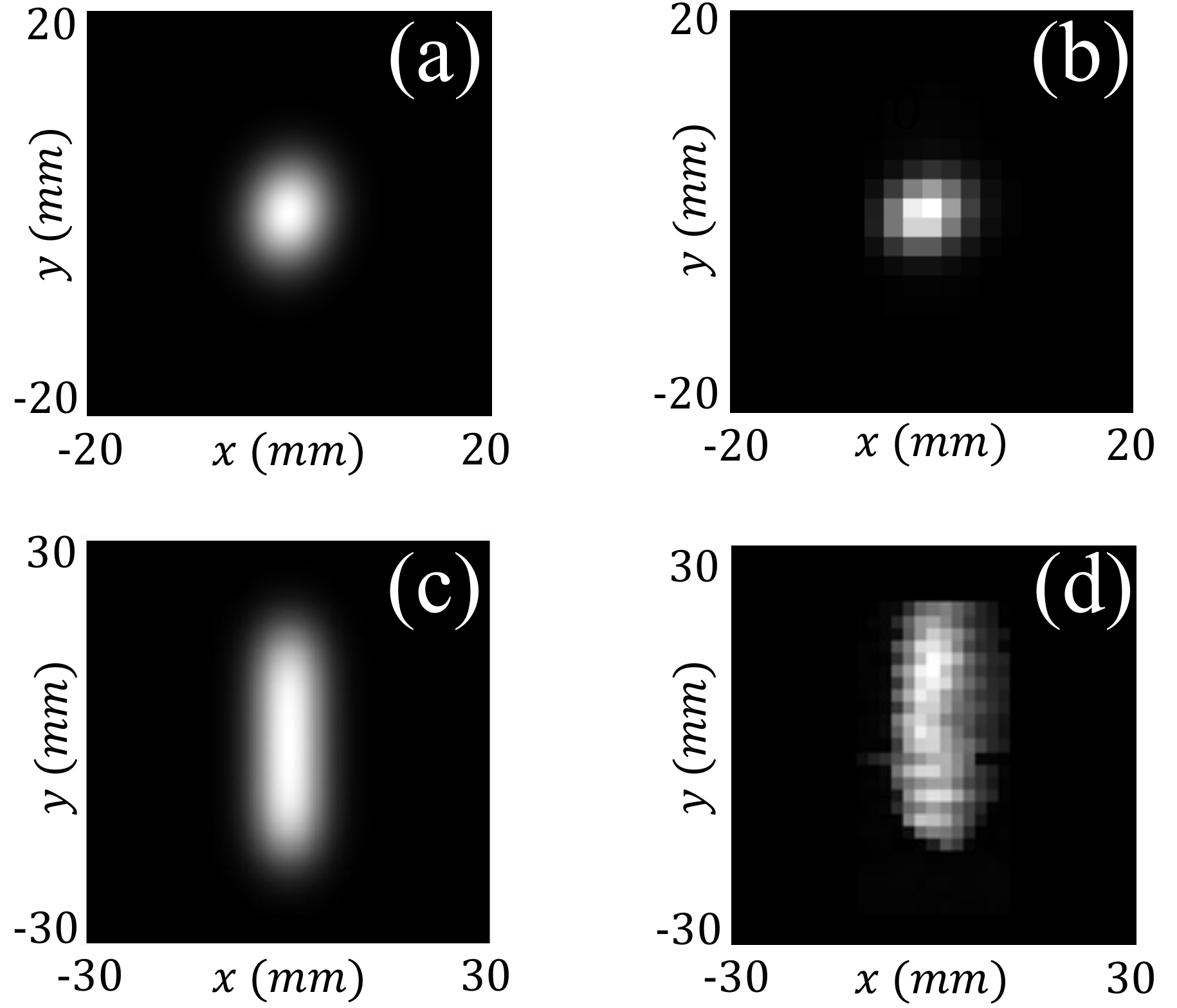}
\caption{(a) Numerical and (b) experimental plots of the conditional spatial distribution of signal photons under loose focusing condition of the pump beam, here $\xi_p \sim$0 i.e. without using any lens. (c) Numerical and (d) experimental plots of the conditional spatial distribution of signal photons under tight focusing condition of the pump beam, here $\xi_p$=0.832 obtained using a lens of focal length $f$=50 mm.}\label{biphotonmodes}
\end{center}
\end{figure}

Now, in order to see how the difference in photon number densities of signal and idler affect the conditional coincidence images of down-converted photons, we recorded the conditional spatial distribution of signal photons under two extreme pump focusing conditions, i.e. loose pump focusing (Fig. \ref{biphotonmodes}(a), (b)) and tight pump focusing (Fig. \ref{biphotonmodes}(c), (d)).
To image conditional biphoton modes, the fiber collimators (FC$_1$ \& FC$_2$) kept in signal and idler arms were mounted on XY translational stages. The fiber collimator in the idler arm was adjusted to get maximum individual counts and fixed. Then the fiber collimator in the signal arm was moved manually along X \& Y directions with a step size of 1 mm. A total of 400 and 900 spatial points were considered in the scanning in loose pump focusing and tight pump focusing cases respectively. The individual as well as the coincidence counts were recorded for each array point. Figures \ref{biphotonmodes} (a), (b) and \ref{biphotonmodes} (c), (d) show the numerical and the corresponding experimental results for the conditional coincidence imaging under both the conditions. The numerical results are obtained by plotting the density plots of normalized conditional signal mode function described by the Eqn. (\ref{condmodefn}), for loose and tight pump focusing conditions respectively. It is clear from the figure that the experimental results are in good agreement with the numerical simulations. We also observe that the overlap extent of conditional signal and idler modes decreases under tight focusing condition as one of the conditional modes becomes elliptic for a given idler coupled in a single mode fiber.

\begin{figure}[h]
\begin{center}
\includegraphics[width=1\textwidth]{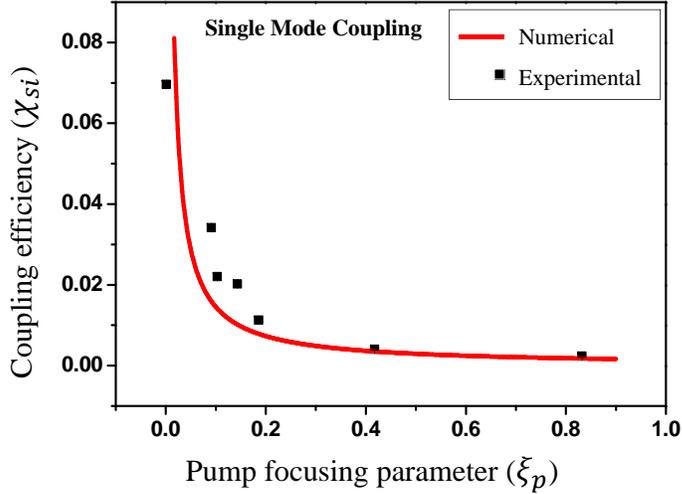}
\caption{(Colour online) Experimental plot of the variation of photon pair collection efficiency with respect to pump focusing parameter. Error bars have been subsumed by the thickness of experimental points.}\label{singlemodefocusexptfa}
\end{center}
\end{figure}

In observing the effect of pump focusing on biphoton modes, we tried to quantify the degree of overlap between conditional signal and idler modes in down-conversion and its variation with pump focusing parameter. For this, we calculated the biphoton mode coupling efficiency ($\chi_{si}$) for different pump beam focusing parameters. Figure \ref{singlemodefocusexptfa} shows the variation of experimentally obtained $\chi_{si}$ with the pump beam focusing parameter along with the numerical results. 
We achieved the maximum coupling efficiency of only ~8$\%$, which is attributed to the mismatch of the numerical aperture of  fiber collimator (FC) and single mode fiber used for the experiment. To calculate the focusing parameter of the pump beam, we calculated the beam diameter at the focus. The diameters of two collection modes projected onto the crystal are calculated as $w_0$=456 $\mu$m. We observed an asymptotic decrease in coupling efficiency with the pump focusing parameter, which matches with the theory of single mode fiber coupling of down-converted photons given in \cite{castelletto}.  From the numerical plot given in Fig. \ref{theoryplot}, it is clear that the influence of collection mode diameter is nominal under tight focusing condition ($\xi_p > 0.1$) whereas it is significant for loose focusing condition ($\xi_p < 0.1$). From the graphs, one can also observe that the coupling efficiency is higher for the loose focusing than the tight focusing, which is also clear from our experimental results given in Fig. \ref{singlemodefocusexptfa}. In our experiment, we used the same crystal length and collection mode diameter in order to study the effect of focusing on the coupling efficiency. We observe that the coupling efficiency mainly depends on the overlap of two conditional modes and the asymmetry present in the ring of down-converted photons.

\section{Conclusion}
We have studied the effect of pump beam focusing on photon pair coupling efficiency of signal and idler in Type I non-collinear spontaneous parametric down conversion. We have experimentally verified that the conditional coupling efficiency of the down-converted modes into a single mode fiber varies asymptotically with the pump beam focusing parameter. This behaviour is attributed to the the asymmetry in the spatial distribution of down-converted photons with the pump beam focusing parameter, due to which the conditional modes of down-converted photons become elliptic. From our observations, we conclude that a loosely focused or almost collimated pump beam inside a thin crystal is the best pre-detection scenario for an efficient fiber coupled source of entangled photon pairs. These mode coupling techniques will be very useful in generating better sources of entangled photons for different quantum information techniques \cite{kok}.

%\bibliography{mybibfile.bib}
%\bibliographystyle{elsarticle-harv.bst}

\end{document}